# Binary Dispersion Formulations of Mid-Infrared Lossless Dielectric Contrast Gratings

K. Araki[1,a)] and R. Z. Zhang,[2]

[1]*School for Engineering of Matter, Transport and Energy, Arizona State University, Tempe, Arizona 85287*

[2]*Department of Mechanical Engineering, University of North Texas, Denton, Texas 76207*

The Zero Contrast Grating (ZCG) is a promising candidate for moded and dispersive mid-infrared (MIR) photonics for its broadband transmittance and reflectance greater than 99.9%. The optimal grating design including refractive index and grating parameters is investigated, and is compared to its analogue the High Contrast Grating (HCG). This study provides insights on how ZCG can manipulate the electromagnetic waves within the low refractive index dielectric bars. The foundations following its near-perfect transmittance and reflectance due to phase matching are described from the perspective of diffraction orders and dispersion relations. Both $0^{th}$ and $1^{st}$ order diffraction contribute to construction or destruction of interferences that occur at the output plane, which is a significant finding in differentiating ZCG and HCG. Applications for the ZCG can be found in selective bandpass windows, beam splitters, broadband reflectors, and narrowband thermal emitters.

**I. INTRODUCTION**

Photonic materials and optical designs for tailoring infrared radiation have applications across spectroscopy, infrared imaging, remote sensing, thermal regulation, medical diagnostics, and more.[1-4] Yet some fundamental optical phenomena are not as elucidated at the feature size boundary between the sub- and at-wavelengths. For instance, resonances that occur in the Fabry-Perot interferometer, plasmonic resonances in electron-rich gratings, and constructive or destructive interference effect are significant in determining the ability to perfectly transmit or reflect or absorb light at resonant conditions.[5] Mid-infrared (MIR) wavelengths applications require high quality electromagnetic wave propagation conditions amid challenges in materials selection and size-dependent sensitivity to micromanufacturing variability.[6]

Among MIR photonics, mirror-like reflectors and wideband bandpass filters are crucial for either reflectance or transmittance over 99.9% for optical and thermal performance enhancement. The well-known optical structure for broad reflectance is established via multilayers consisting of low and high index dielectrics, known as the Distributed Bragg Reflector (DBR).[7] The simple DBR has a history of investigation toward alternating multilayer patterned structure, which utilize Fano resonance as well as Fabry-Perot resonances.[8] The constructive interference at the input plane of the DBR is a mechanism achieved with quarter-wave thickness. With an increase in the number of layers, a DBR can improve the stop band with near-perfect reflectance. In contrast, this photonic band can be created with an optimized grating made of high-index material such as silicon. This new arrangement, called the High Contrast Grating (HCG), was developed to replace DBR in vertical-cavity surface emitting laser (VCSEL).[9, 10] HCG has been nano/micro-fabricated for both one-dimensional (grooves) and two-dimensional (islands) variations.[11] The key factor in HCG was determined to be the destructive interference of $0^{th}$ order transmission at the exit plane.[12] $0^{th}$ order transmission is the non-diffractive near-normal dispersion. A Fabry-Perot round trip (FP-RT) governs the index contrast at the interface from high-to-low such that electromagnetic wave experience round-trip within the high-index bar.[11] Hence, 99.9% reflectance is initiated without the use of multilayers, in which ultra-low emissivity can be accomplished by outperforming bare reflective metal surfaces in thermal insulation.[13]

99.9% transmittance can be obtained using the Zero Contrast Grating (ZCG) due to destruction of index contrast at the exit plane of the grating structure.[14, 15] This index-matching interface between the lossless dielectric grating bars and substrate that serves as the waveguide, can provide simultaneous broadband transmittance and narrowband reflectance or absorptance. The ZCG acts as a guided-mode resonance (GMR) filter.[16] With a GMR, narrowband absorption with high quality factor is tunable by modifying the grating refractive index.[17] Some have utilized ZCG coupling with graphene's plasmonic resonance to tune light transmission, reflection, and absorption.[18] A switchable MIR thermal emitter using thermochromic metal-to-insulating phase transition vanadium dioxide ($VO_2$) or electrochromic tungsten trioxide ($WO_3$) was also shaped by the ZCG design and mechanism.[19] These resonances or interferences are highly dependent on its geometric parameters in relation to the wavelength of the light. The optical performance is often determined by the wavelength-to-period ratio and height-to-period ratio such that the grating structure is optimized for its purpose. In this comprehensive yet succinct analysis, we provide an in-depth study of ZCG's design, modalities, and comparison to the HCG. Due to its dispersion regimes that are dependent on several geometric values and light characteristics, ZCG can be a more suitable single material thin film substance for bandpass filtering, dichroic filtering or reflectors, and narrowband absorbers.[14, 19-23]

## II. RESULTS AND DISCUSSION

This work investigates mechanisms behind the broadband high transmission using ZCG aimed at the mid-infrared (MIR) region. The optimal 1D grating is introduced which produces transmittance over 99%. Here, the normal transmittance is focused on transverse magnetic (TM) wave calculated using the rigorous-coupled wave analysis (RCWA).[24] The refractive index, grating height, grating period, filling ratio, and wavelength are denoted as $n$, $h$, $\Lambda$, $\phi$ and $\lambda$, respectively, as shown in Fig. 1. These parameters can be optimized, which consist of high transmittance as can be evaluated with the integrated transmittance only greater than 99%. The integrated surface transmittance can be written in two ways in respect with (a) grating height-to-period ratio $h' = h/\Lambda$ and filling ratio $\phi$, and (b) refractive index $n$ and wavelength-to-period ratio $\lambda' = \lambda/\Lambda$ which is expressed as,

$$S_\tau(h',\phi) = \iint_S \tau d\lambda' dn \tag{1a}$$

$$S_\tau(\lambda',n) = \iint_S \tau dh' d\phi \tag{1b}$$

Here, the refractive index is limited to low index material such as potassium bromide (KBr) ($n = 1.4$) to high index material germanium (Ge) ($n = 4.0$). Silicon (Si) and Ge have been good candidates for HCG for their high refractive index and low absorption in the infrared region.[25, 26] In comparison to integrated surface transmittance of ZCG shown in Fig, 1 (c) and (e), the optimal grating structure for HCG in terms of surface reflectance $S_\rho$ is evaluated for reflectance greater than 99% as demonstrated in Fig. 1 (d) and (f). Note that the contour bar is normalized to maximum integrated values. As a consequence of the integration, Eq. (1.1) determines which grating structure can be applied to wide range of refractive index and wavelength. Hence, the optimal grating height-to-period ratio and filling ratio is $h/\Lambda = 0.46$ and $\phi = 0.52$ respectively for ZCG, resulting in lossless grating bars with near-square cross-section in length of the half the period. On the other hand, HCG reveals optimum at $h/\Lambda = 0.66$ and $\phi = 0.64$ respectively, which closely matches with previous report.[12] Thus, at the optimal grating height and filling ratio, ZCG can obtain transmittance greater than 99% in any lossless material from $n = 1.4$ to $n = 4.0$ as shown in Fig. 2 (a). In comparison, HCG can only provide reflectance greater than 99% for refractive index larger than 2.0 but within narrow wavelength ranges. In contrast to the integrated transmittance over wavelength and refractive index, Eq. (1.2) discovers which refractive index has the highest flexibility to the grating parameters. As can be indicated from Fig. 1 (e) and (f), the lowest refractive index has no dependency on the grating structure due to weak interference effects at both input and output planes of the grating, while HCG is highly sensitive to index ($n>2.0$) and wavelength ($\lambda/\Lambda>2.0$) which relies on an interference effect. Similar results have been investigated for monolithic HCG.[27, 28] Therefore, HCG has limited variation in grating structure, index, and spectral band in contrast to ZCG for higher grade of transmittance or reflectance. This trade-off between the two gratings also applies to diffraction patterns.

Diffraction orders are an important and distinctive characteristic of the periodic structure since the light path can be manipulated with the periodicity of the grating and the wavelength of the incident light. The use of diffraction orders has been investigated recently for unconventional redirection of light, often called metagratings.[29, 30] Fig. 3 demonstrates how ZCG and HCG have dissimilarities in $0^{th}$ order and $1^{st}$ order transmittance and reflectance, respectively. This figure exhibits that ZCG becomes highly dependent on the $1^{st}$ order diffraction for higher index while HCG have no contribution in its $1^{st}$ order due to the cancellation of $0^{th}$ order transmission within the same wavelength range. With an additional solid layer under the lossless grating bars, the thin film disturbs the Fabry-Perot round trip (FP-RT) in the ZCG, leading to weak $0^{th}$ order reflectance. Hence, phase mismatching occurs at the input plane of the grating while allowing $1^{st}$ order transmission through the grating. In comparison, the HCG is recognized for its reflectance greater than 99.9% because it experiences full FP-RT to construct strong interferences at the input plane. The HCG's $1^{st}$ order reflectance is completely suppressed as shown in Fig. 3 (d). This reflection is governed by the FP-RT matrix where two eigenmodes (i.e., single and even waveguide modes) play a role in broadband high reflectance when the grating period $\Lambda$ is comparable to incident wavelength $\lambda$.[31] Thus, when two modes intersect with each other, which meets FP conditions, the reflectance vanishes as can be seen in the dark regions ($\rho = 0$) in Fig. 3 (c) due to phase difference of dual modes by multiples of $\pi$.[31] Similar phenomena take place in ZCG in the $1^{st}$ order transmission, where 99.9% transmission is achieved due to excitation of two FP-RT modes, as can be indicated in Fig. 4. In other words, single and even waveguide modes cancel out the $0^{th}$ order transmission, and what remains is the $1^{st}$ order transmission greater than 99.9%. Therefore, one can split the incoming beam using ZCG between shorter wavelength regions (UV/visible) and longer wavelength regions (far-IR) such that only MIR radiation can be transmitted normally down the film. The $0^{th}$-to-$1^{st}$ order cutoff wavelength can be determined from the longitudinal wavevector being zero, as a function of the grating period. The primary distinguishment of ZCG and HCG is diffraction order splitting due to differences in their dispersions and phase matching conditions.

While HCG is a high-index periodic bar surrounded by air or vacuum ($n = 1.0$), ZCG has no index gradient at the output plane of the bars. Both HCG and ZCG bring uniqueness in propagation constants which represent the destructive interference of reflection. The dispersion relation of HCG is determined by,

$$-k_{air} \tan\left(\frac{k_{air}(1-\phi)\Lambda}{2}\right) = n_{bar}^{-2} k_{bar} \tan\left(\frac{k_{bar}\phi\Lambda}{2}\right) \tag{2}$$

where $k_{air}$ and $k_{bar}$ is the wavevector along x-axis and $n_{bar}$ is the refractive index of the grating bars.[31] The longitudinal wavevector $\beta$ can be calculated from the relation, $\beta = \sqrt{(2\pi/\lambda)^2 - k_{air}^2}$. Thus, the grating height $h$ for HCG for constructive interference is obtained from phase matching as,

$$2\psi = 2\beta h = 2m\pi \tag{3}$$

where $m = 0, 1, 2, \ldots$ such that $h = m\pi\beta^{-1}$. The obtained dispersion curve is overlayered in white dashed lines on the reflectance contour of standalone HCG with filling ratio of $\phi = 0.64$ and $n_{bar} = 3.4$ as shown in Fig. 5 (a). The black vertical lines indicate cut-off wavelengths where two FP-RT eigenmodes excite (single and even waveguide modes) within the high index bars. Thus, HCG can provide broad reflectance boxed with reflectance greater than 99.9% achieved with the two eigenmodes. On the other hand, ZCG with a semi-infinite substrate experiences additional phases that need to be considered for phase matching conditions between (a) the grating to air, and (b) the grating to its substrate, so that constructive interference between reflected wave has a phase difference of $2m\pi$.[32] In this case, the phase match condition is,

$$2\psi = 2\beta h + \arg(r_{inc}) + \arg(r_{sub}) = 2m\pi \tag{4}$$

where $r_{inc}$ and $r_{sub}$ are the reflectance coefficient between (a) the grating to the air, and (b) grating to the substrate, which is expressed as,

$$r_{inc} = \frac{k_{inc,z}/n_{air}^2 - \beta/n_{eff}^2}{k_{inc,z}/n_{air}^2 + \beta/n_{eff}^2} \tag{5a}$$

$$r_{sub} = \frac{\beta/n_{eff}^2 - k_{inc,z}/n_{air}^2}{\beta/n_{eff}^2 + k_{inc,z}/n_{air}^2} \tag{5b}$$

where $k_{inc,z}$ and $k_{sub,z}$ is the longitudinal wavevector in incident and substrate region respectively.[33] Here, $n_{eff}$ is the effective refractive index of the grating region determined by $n_{eff}^2 = n_{bar}^2 n_{air}^2 / \left[ \phi n_{air}^2 + (1-\phi) n_{bar}^2 \right]$. By converting the phase angles, Eq. (4) can be rewritten in terms of grating height $h$ expressed as follows.[16, 33] The following relationship can also be determined from dispersion relation of a guided mode resonator (GMR)[34] by matching the refractive index of the waveguide layer and the grating bar, but also taking a semi-infinite substrate.

$$h = \left[ m\pi + 2\tan^{-1}\left( \frac{k_{inc,z} n_{eff}^2}{\beta n_{air}^2} \right) + 2\tan^{-1}\left( \frac{k_{sub,z} n_{eff}^2}{\beta n_{sub}^2} \right) \right] \beta^{-1} \qquad (6)$$

In consequence, ZCG with semi-infinite substrate follows the dispersion curve shown in Fig. 5(d), which exhibits low reflectance due to FP-RT interruption at interface of grating bar and substrate. Accordingly, broader transmittance can be obtained with ZCG which offers beam splitting between 0$^{th}$ and 1$^{st}$ order for shorter and longer wavelength regions. For a finite thin layer underneath the grating bars, the reflectance remains high, as shown in Fig. 5(b) and (c) for the case of substrate thickness of 0.2 μm and 0.4 μm, respectively. Both indicate that reflectance band structure shifts from HCG to ZCG with increase in substrate thickness. In other words, the destructive interference curves fall between HCG and ZCG dispersion curves, eventually merging with ZCG curve for semi-infinite thickness where FP-RT condition for reflectance greater than 99.9% is destroyed. On the strength of this observation, ZCG can be applied as a broadband mirror by stacking it on a metallic substrate since waveguide modes are weak, improving width of wavelength ranges with high reflectance compared to HCG itself.[14]

Considering physical manifestations of ZCG for MIR devices and applications, ZCG has an advantage over HCG in optical filtering performance as well as mitigates more complex manufacturing process, where ZCG is supported by a substrate.[35] Whereas the comparably configured HCG requires a suspended and enclosed thin film.[36] ZCG also offers good performance composed of lower index materials, which may be easier to etch compared to harder high-index crystals.[37] In replacement of gradient-index metasurfaces, such as cones, pyramids, and other inverted diffraction-range geometrical microstructures,[38, 39] the ZCG enhances higher grade transmission and reduces reflection by multiple orders. We note both one-dimensional and two-dimensional patterns are suitable for ZCG as optical filters. Widely available and established fabrication methods, such as electron beam lithography, laser ablation, deep dry etching, and nanoimprinting,[40, 41] can be used to cast wide-area and supported ZCG optical devices.

## III. CONCLUSION

In conclusion, we demonstrated the comparable and differential mechanisms behind the transmittance and reflectance over 99.9% for ZCG and HCG. The dispersion surface integral of transmittance and reflectance provides insight into the optimal ZCG and HCG structure for their best performance. The best ZCG has a filling ratio of 0.52 and grating height-to-period ratio of 0.46, while HCG has a 0.66 filling ratio and a 0.64 height ratio. Both ZCG and HCG have preference over square shape grating bars, but the major difference in preference falls into index. HCG presents the best reflectance performance over high index material such as silicon and germanium, whereas ZCG is less sensitive to medium refractive index, and both low and high index materials can be applied depending on which wavelength region ZCG is utilized. Most significantly, ZCG controls the splitting of incoming electromagnetic waves into $0^{th}$ and $1^{st}$ diffraction orders, which both anticipating short-band and long-band transmittance over 90%. This splitting is because ZCG has strong destructive interference which is converted to 99.9% transmittance with broader bandwidth, while HCG has strong constructive interference of reflected waves. The binary dispersion relations validate the observation of interference effects on both ZCG and HCG. To this end, ZCG can be an effective beam splitter, metalens, bandpass filter, wideband reflector for infrared or thermal photonics.

**Acknowledgements**

R.Z. is grateful for the research sponsorship by the Army Research Office Grant No. W911NF-23-1-0165. Views and conclusions contained in this manuscript are those of the authors and should not be interpreted as representing the official policies, either expressed or implied, of the Air Force Office of Scientific Research or the U.S. Government.

**Conflict of Interest**

The authors have no conflicts to disclose.

**Data Availability**

The data that support the findings of this study are available from the corresponding author upon reasonable request.

---

[a] Corresponding author. Electronic mail: karaki1@asu.edu

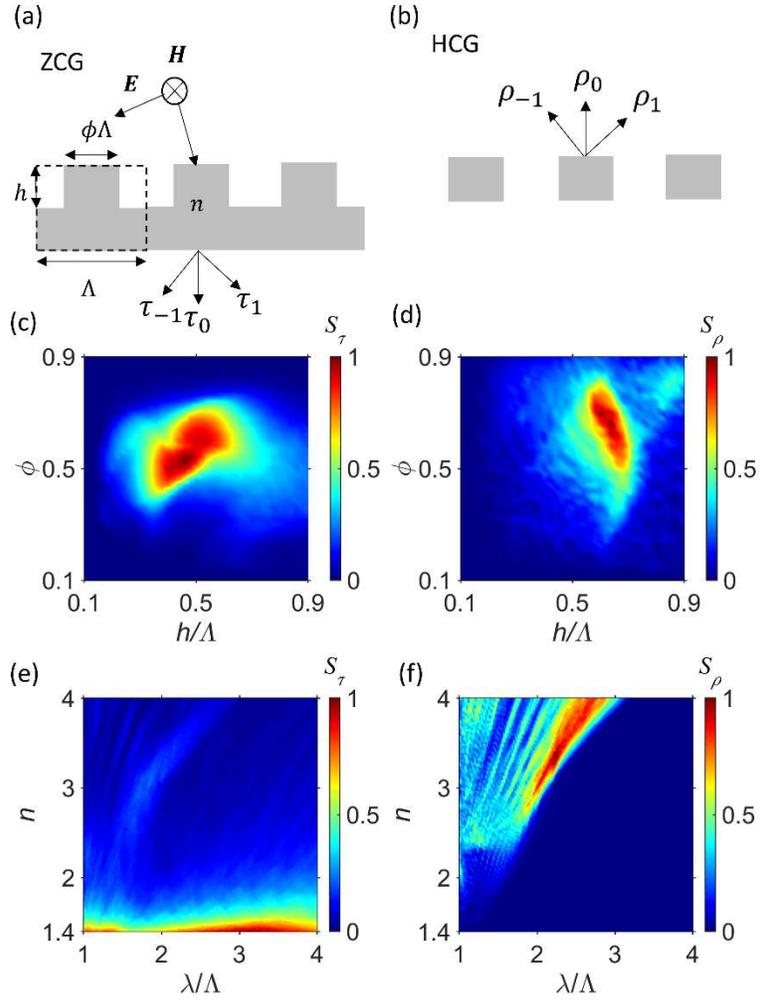

FIG. 1. Nomenclature of (a) ZCG and (b) HCG where $\tau$ and $\rho$ represents transmittance and reflectance respectively. Here, 0,1, -1 are diffraction orders. Surface transmittance integrated in terms of grating height $h$ and filling ratio $\phi$ for (c) ZCG and (d) HCG, and surface reflectance integrated in terms of wavelength $\lambda$ and refractive index $n$ for (e) ZCG and (f) HCG.
10

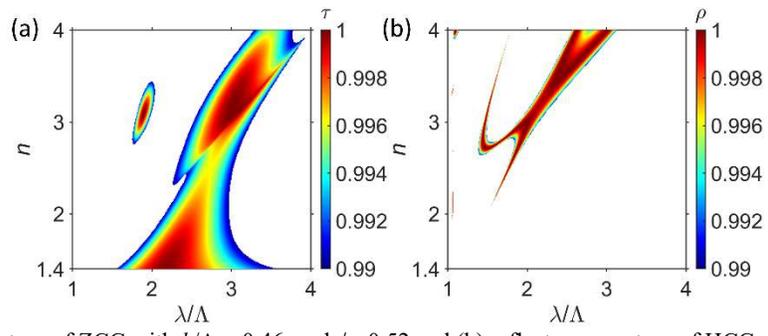
FIG. 2. (a) Transmittance contour of ZCG with $h/\Lambda$ = 0.46, and $\phi$ =0.52 and (b) reflectance contour of HCG with $h/\Lambda$ = 0.66, and $\phi$ =0.64 with respect to refractive index and wavelength.

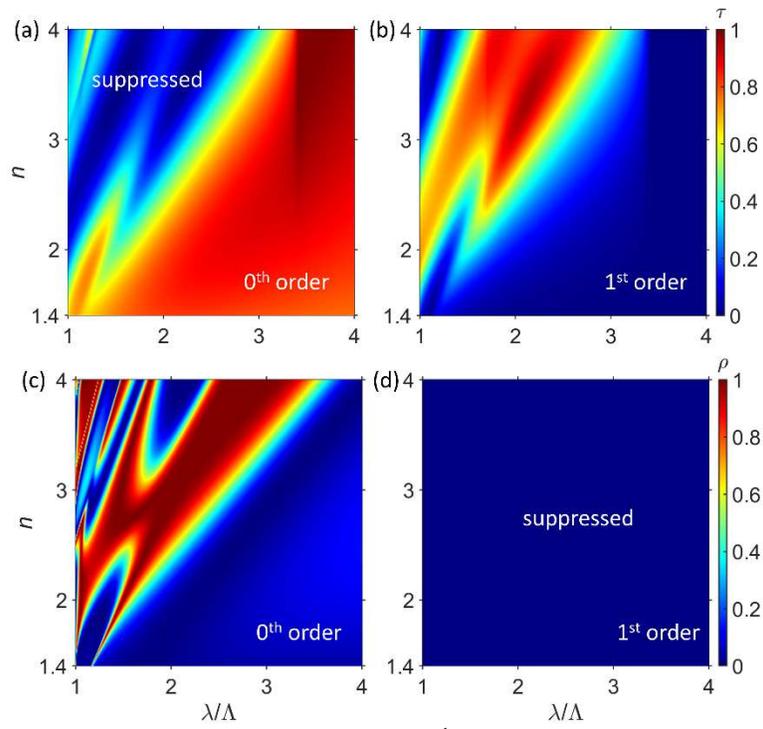

FIG. 3. Transmittance contour of ZCG at $h/\Lambda = 0.46$ and $\phi = 0.52$ for (a) 0$^{th}$ order and (b) 1$^{st}$ order. Reflectance contour of HCG at $h/\Lambda = 0.66$ and $\phi = 0.64$ for (c) 0$^{th}$ order and (d) 1$^{st}$ order.

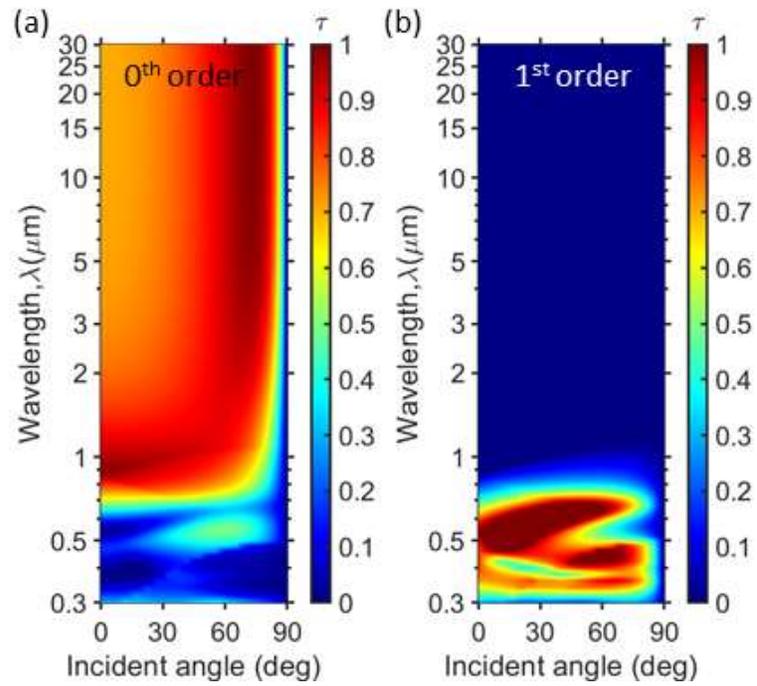
FIG. 4. (a) $0^{th}$ order and (b) $1^{st}$ order transmittance contour of ZCG with $h/\Lambda = 0.46$ and $\phi = 0.52$ where period $\Lambda = 0.25$ μm.

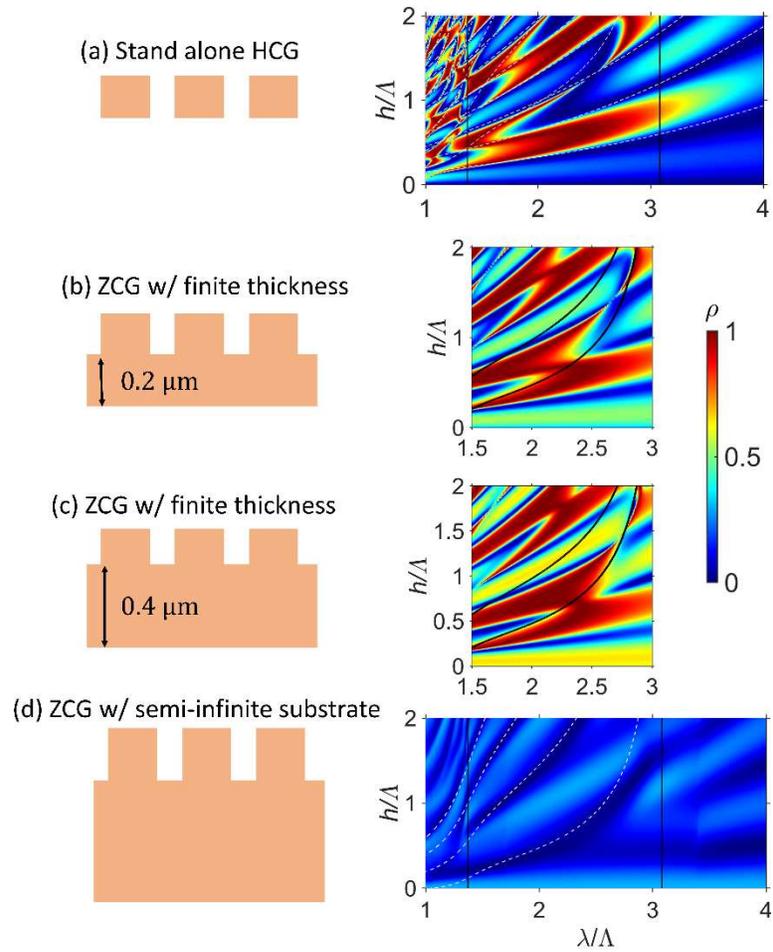

FIG. 5. Reflectance contour of (a) standalone HCG, (b) ZCG w/ finite thickness (0.2 μm), (c) ZCG with finite thickness (0.4 μm), and (d) ZCG with semi-infinite substrate with filling ratio of $\phi = 0.64$ and refractive index of $n_{bar} = n_{sub} = 3.4$. White dashed lines represented in (a) and (d) are dispersion curves for HCG and ZCG respectively. Black solid lines shown in (b) and (c) are dispersion curves for HCG and ZCG.